\documentclass{PoS}
\usepackage{latexsym}
\usepackage{graphicx}
\usepackage{epsfig}

\newcommand{\be}{\begin{equation}}

\newcommand{\ee}{\end{equation}}
\newcommand{\bea}{\begin{eqnarray}}
\newcommand{\eea}{\end{eqnarray}}
\newcommand{\bef}{\begin{figure}}
\newcommand{\eef}{\end{figure}}
\newcommand{\bce}{\begin{center}}
\newcommand{\ece}{\end{center}}

\def\NP{{ Nucl.\ Phys.\ }}
\def\PL{{ Phys.\ Lett.\ }}
\def\PR{{ Phys.\ Rev.\ }}

\def\PRL{{ Phys.\ Rev.\ Lett.\ }}

\def\EP{{ Europ.\ Phys.\ J.\ C}}

\title{Universal strangeness production and size fluctuactions in small and large systems}

\ShortTitle{Universality of strangeness production}

\author{\speaker{P. Castorina}\\
        Dipartimento Fisica ed Astronomia , Universita' di Catania, Italy and INFN ,Sezione Catania,Italy\\
        E-mail: \email{paolo.castorina@ct.infn.it}}

\author{M. Floris\\
       CERN, Geneva, Switzerlan\\
       E-mail: \email{Michele.Floris@cern.ch}}

\author{S. Plumari\\
        Dipartimento Fisica ed Astronomia, Universita' di Catania, Italy and Laboratori Nazionali del
Sud, INFN-LNS, Catania, Italy \\
       E-mail: \email{salvatore.plumari@hotmail.it}}

\author{H.Satz\\
        Fakult\"at f\"ur Physik, Universit\"at Bielefeld, Germany\\
       E-mail: \email{satz@physik.uni-bielefeld.de}}

\abstract{Strangeness production in high multiplicity events gives indications on the transverse size fluctuactions in nucleus-nucleus ($AA$), proton-nucleus ($pA$)  and proton-proton ($pp$) collisions. In particular the  behavior of strange particle hadronization in "small" ($pp,pA$) and "large" ($AA$) initial configurations of the collision  can be tested for the specific particle species, for different centralities and for large fluctuations of the transverse size in $pA$ and $pp$ by using the recent ALICE data. A universality of strange hadron production emerges by introducing a dynamical variable proportional to the initial parton density in the transverse plane.}  

\FullConference{EPS-HEP 2017, European Physical Society conference on High Energy Physics\\
		5-12 July 2017\\
		Venice, Italy}

\begin{document}

\section{Introduction}
Recent experimental results in $pp$ and $pPb$ collisions at the Large Hadron Collider energies  show a strong similarity to those observed in $AA$ collisions, where the formation of a Quark-Gluon Plasma (QGP) is expected. In particular, ALICE collaboration reported \cite{Alice1} the enhanced production of multi-strange hadrons in high energy, high multiplicity, $pp$ events, previously observed in $Pb-Pb$ collisions. These results, suggested on theoretical grounds in refs. \cite{noi1, noi2}, indicate that the final system created in  "small" initial settings ,i.e. $pp$ and $pA$, in high energy collisions is essentially the same as the one produced in "large" initial nucleus-nucleus configurations.

In this note, we show that a universal behavior of strangeness production in small and large systems emerges  by introducing  a specific dynamical variable, which takes into account the transverse size of the initial configuration and its fluctuactions in high multiplicity events.

The starting point is the study of hadron production in high energy collisions by the statistical hadronization model (SHM): the relative hadron multiplicities are obtained by considering an ideal gas of hadrons and resonances at a temperature $T$ and baryonchemical potential $\mu$ \cite{becaT, PBM}. However, the SHM overpredicts the strange hadron production in $e^+ e^-$, $pp$  collisions up to RHIC energy and in nuclear collisions up to SPS energy. Therefore an ad hoc parameter $0< \gamma_s \le 1$ has been introduced \cite{LRT} in such a way that the production of a hadron with $n$ strange quarks/antiquarks is suppressed by a factor $\gamma_s^n$.
In nuclear collisions at high energy $\gamma_s = 1$ and, in this sense, such
 collisions show a strangeness enhancement with respect to elementary 
($pp,pA$,$e^+e^-$) ones, where up to RHIC energy, $\gamma_s$  is less than unity. 

In Fig.1   
the behavior of $\gamma_s$ as function of the collision energy $\sqrt s$ 
in $pp$, $pPb$ and heavy ion ($Pb-Pb$, $Au-Au$, $Cu-Cu$) collisions is depicted \cite{data}.


\begin{figure}
 \begin{minipage}[b]{6.0cm}
   \centering
   \includegraphics[width=5.5cm]{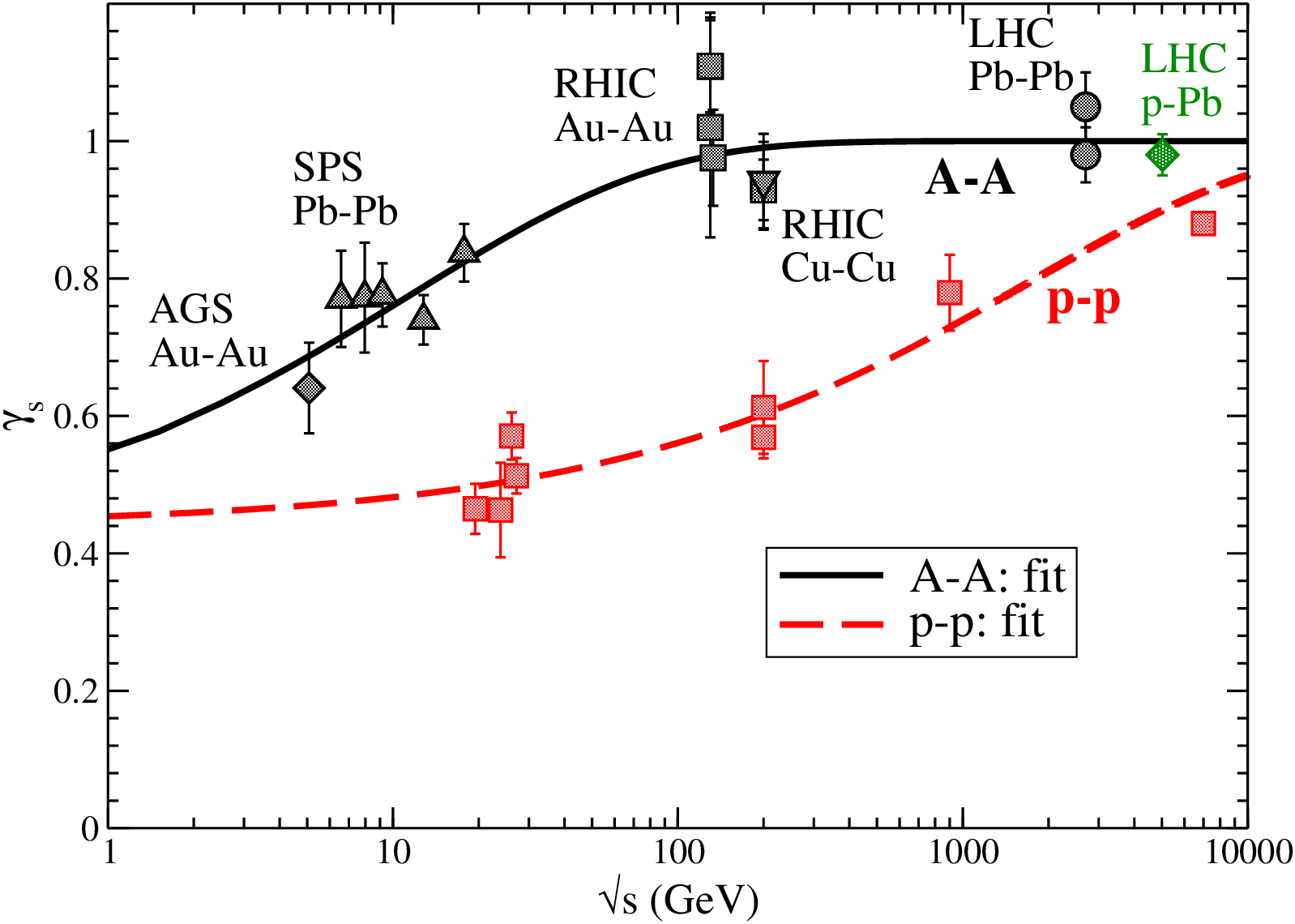}
   \caption{The strangeness suppression factor $\gamma_s$ as function of the
collision energy $\sqrt s$ for $pp$ (red symbols), $Pb-Pb$, $Au-Au$, 
$Cu-Cu$ (black symbols) and $p-Pb$ (green symbol) collisions.}
 \end{minipage}
 \ \hspace{2mm} \hspace{3mm} \
 \begin{minipage}[b]{6.0cm}
  \centering
   \includegraphics[width=5.5cm]{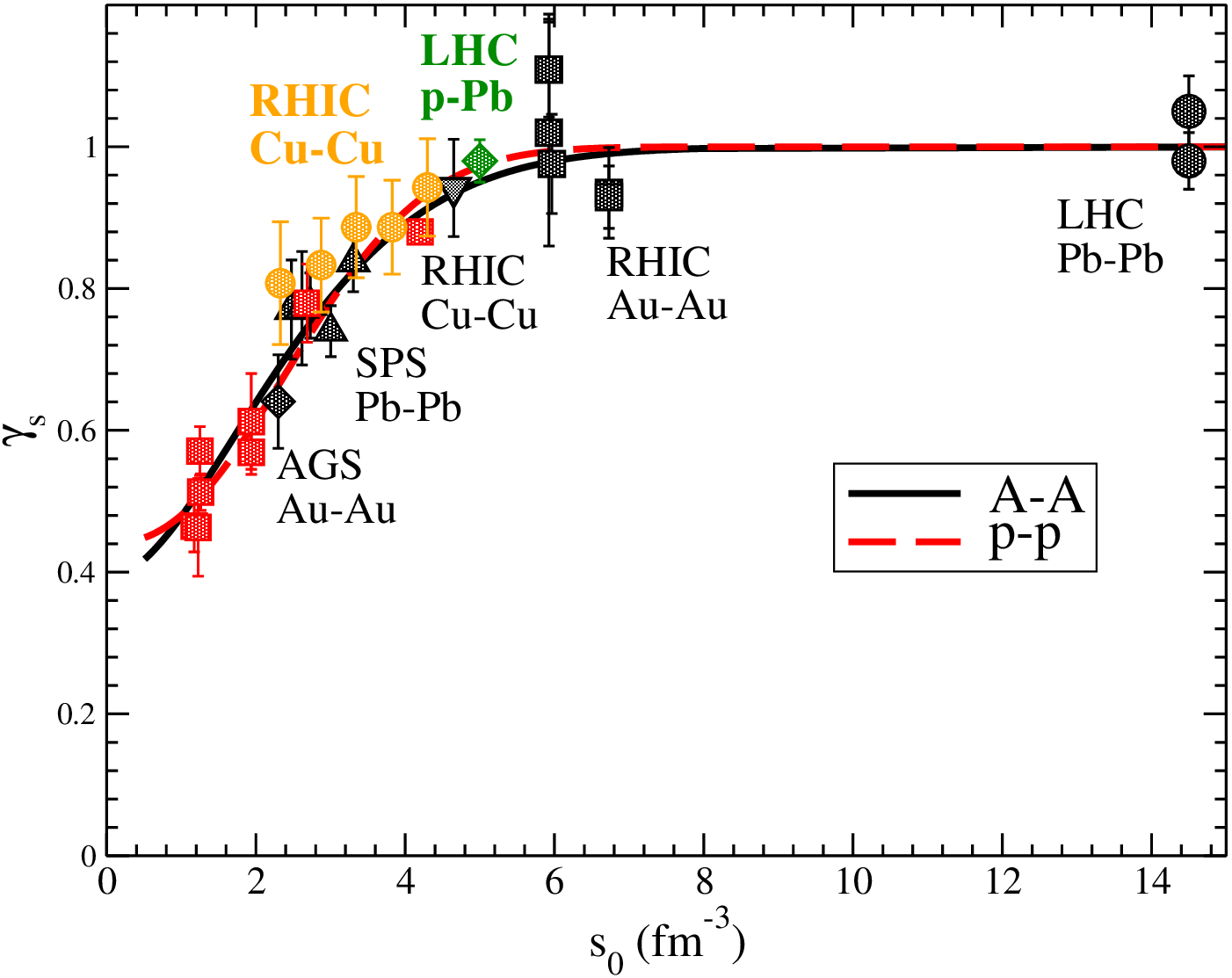}
   \caption{The strangeness suppression factor $\gamma_s$ as function of the
initial entropy density for the same points in Fig.1. The yellow points are for $Cu-Cu$ at different centrality , see \cite{noi2}.}
 \end{minipage}
\end{figure}
The curves in Fig.1 are interpolating fits, given in \cite{data}:
\be
\gamma_s^A(s) = 1 - a_A \exp{(-b_A \sqrt{A \sqrt s})},
\label{10}
\ee
for $AA$ and
\be
\gamma_s^p(s) = 1 - a_p \exp{(-b_p s^{1/4}}),
\label{11}
\ee
for $pp$, with $a_A=0.606,~a_p=0.5595,~b_A=0.0209,~b_p=0.0242$. In the next sections we shall show explicitely that a universal behavior of strangeness production can be derived if one takes into account not only the energy  but also the transverse size of the colliding system.

\section{Universality of strangeness production}
In the 1D Bjorken hydrodynamical model \cite{bjm} the initial entropy density of the collisions, $s_0$, at the termalization time $\tau_0$, is given by
\be
s_0 \tau_0 \simeq \frac{1.5 A^x}{\pi R^2_x} (\frac{dN_{ch}}{dy})_{y=0}^x
\ee
where $x$ indicates $pp,pA,AA$, $R$ is the system radius, $N_{ch}$  the charge multiplicity and $y$  the rapididity. 
On the other hand the behavior of  $dN_{ch}/dy$ as a function of $\sqrt{s}$ is known for different targets and it is given by \cite{mult} ($N_{part}$ is the number of partecipants)
\be
\frac{2}{N_{part}} (\frac{dN}{dy})_{y=0}^{AA} = a(\sqrt s)^{0.3} + b,
\ee
with $a=0.7613$ and $b= 0.0534$ for $AA$ collisions and
\be
\left({dN \over dy}\right)_{y=0}^{pp} = a(\sqrt s)^{0.22} + b,
\label{9}
\ee
where $a=0.797$ and $b= 0.04123$ for $pp$.

By previous eqs.(1.1,1.2,2.1,2.2,2.3) one can eliminate $s$ to obtain  the
suppression factor $\gamma_s$ as a function of the initial entropy
density. The result is shown in Fig.2: all the points in Fig.1 are now on a
universal curve and any difference among $pp,pA,AA$ has disappeared
\cite{CPS}. Fig.2 has been obtained by considering $\tau_0 \simeq 1$ fm and the following transverse radii (associated with the average multiplicity):  $R_A = 1.25 A^{1/3}$  fm for nuclei, $R_p = 0.8$ fm for proton and $R_T = R_p (0.5 {\bar N}_{\rm part})^{1/3}$ for $pA$ system, with ${\bar N}_{\rm part}\simeq 8$ \cite{Abelev} for $pPb$. The entropy density, $s_0$, is directly related with the saturation scale in the Color Glass Condensate (CGC) model \cite{larrylecture}, $Q^2_s \simeq (dN/dy)/A_T$ ($A_T$ being the transverse area) and, indeed, a plot of $\gamma_s$ as a function of $Q^2_s$ gives a universal curve similar to Fig.2. More generally, the universal behavior emerges if a dynamical variable proportional to the parton density in the transverse plane is considered.    
In the next section we shall verify the  universal behavior by comparison with ALICE data on multi-strange hadron production in high multiplicity events.

\section{Strangeness production and size fluctuactions}

The description of hadron final states in $pp$ at high energy and mutiplicity requires a more refined analysis of the transverse area in the collisions. Indeed, the previous values of the transverse radii refer to  average multiplicities but large fluctuactions of the transverse size are obtained, for example, in the (CGC) \cite{trans-area} and are crucial to reproduce the experimental data at high multiplicity.

For nucleus-nucleus collisions we take into account the fluctuactions of the
transverse area by a rather standard approach, i.e. the Glauber Montecarlo calculations in ref. \cite{glau}; 
however, for $pp$ and $pA$ one needs a specific model. Theoretical and phenomenological analyses carried out in the CGC model give, for $pp$ and $pA$, the relation between the transverse area and the multiplicity reported in Fig.3 \cite{trans-area}.
\begin{figure}[h]
\centerline{\psfig{file=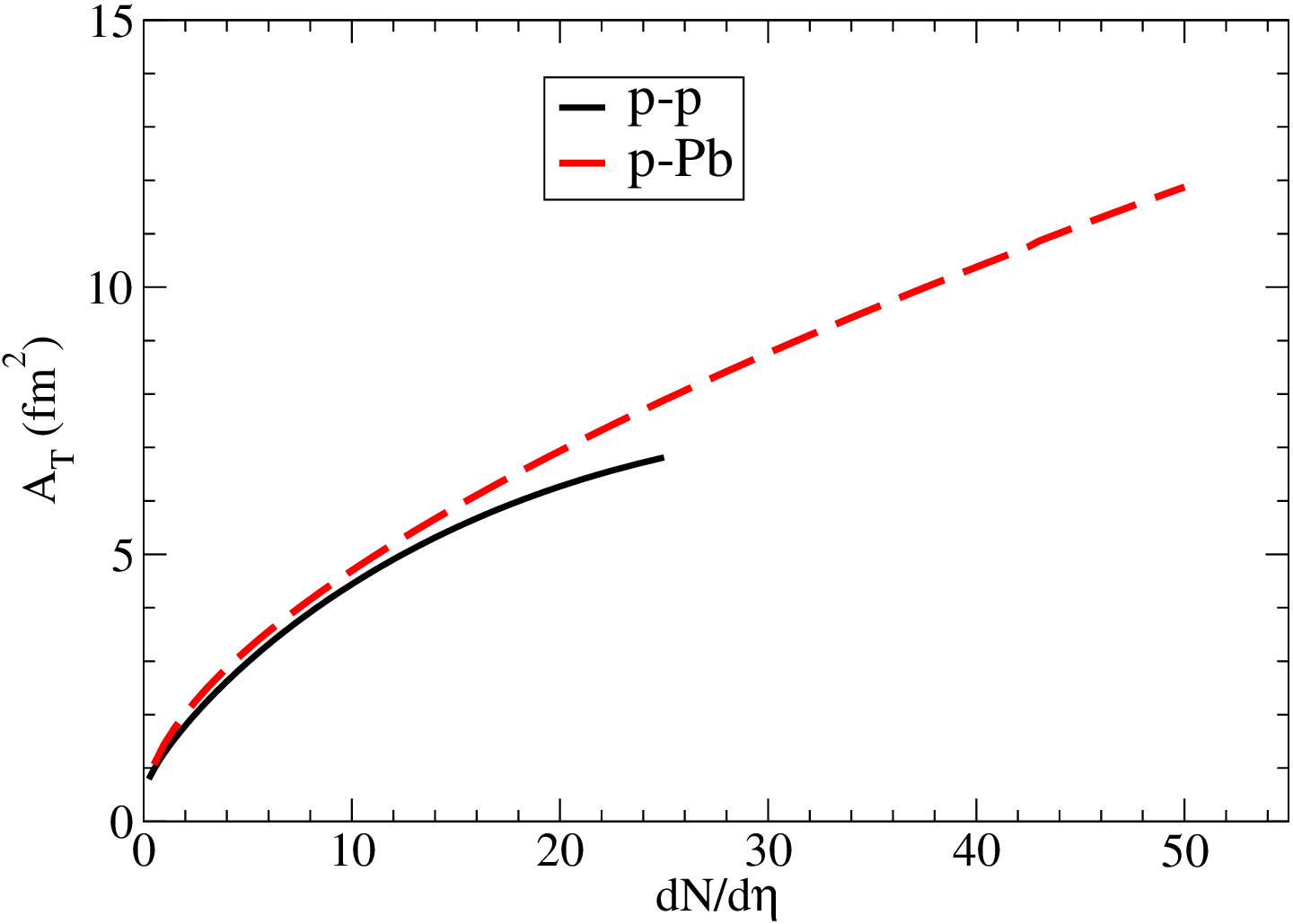,width=5cm}}
\caption{The relation between transverse area and multiplicity obtained in refs. \cite{trans-area}}
\end{figure}
 By using these numerical results  and repeating the analysis of Sec.2,  one gets that the data for multi-strange hadron production for $AA$ and for $pp$ and $pA$ at high multiplicity stay on universal curves, 
for different species, as shown in Fig.4. In Fig.5 we show the comparison of ALICE data for different species with the corresponding $\gamma_s^n$ suppression factor. In conclusion the strangenesss production in $AA,pp$ and $pA$ in high energy collisions suggests a universal behavior which emerges by taking into account the transverse size of the systems and its fluctuactions in high multiplicity events. However, in events with large fluctuactions of the gluon field in the initial configuration the transverse size of small systems is model dependent \cite{models} and therefore a more complete analysis, which takes into account ALICE data at higher energy \cite{eps}, is in progress.   
\begin{figure}
 \begin{minipage}[b]{5.5cm}
   \centering
   \includegraphics[width=5.0cm]{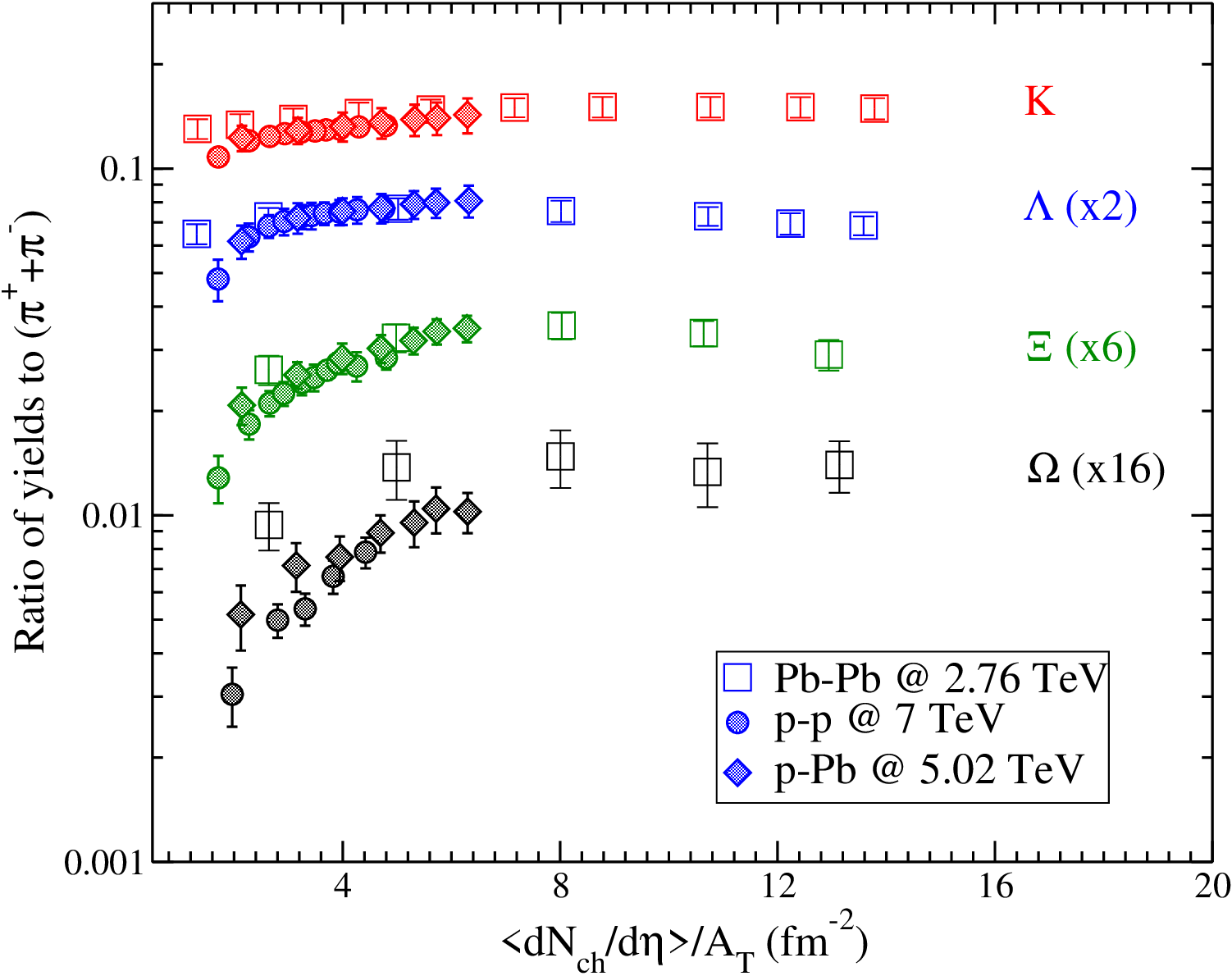}
   \caption{Universality of strangeness production for different species of strange hadrons and in high multiplicity events.}
 \end{minipage}
 \ \hspace{2mm} \hspace{3mm} \
 \begin{minipage}[b]{5.5cm}
  \centering
   \includegraphics[width=5.0cm]{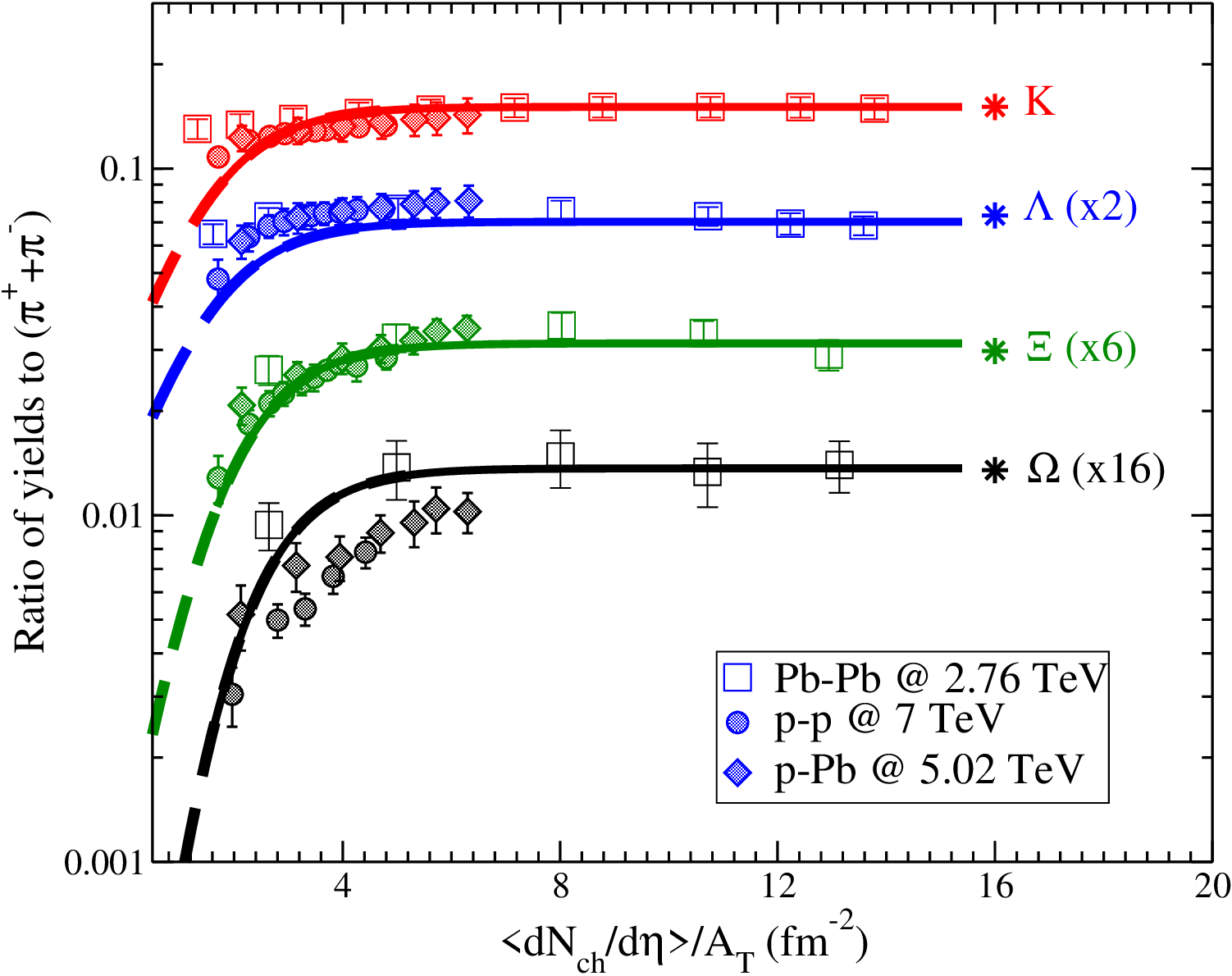}
   \caption{Comparison of ALICE data with the corresponding $\gamma_s^n$ suppression for different hadrons. The stars indicate the SHM result with $\gamma_s=1$.}
 \end{minipage}
\end{figure}

\end{document}